\documentstyle[aps,epsfig]{revtex}

\begin{document}

\def\Pom{{\bf I\!P}}
\def\be{\begin{equation}}
\def\ee{\end{equation}}
\def\bea{\begin{eqnarray}}
\def\eea{\end{eqnarray}}
\def\pt2{p^2_\perp}
\def\lsim{\mathrel{\rlap{\lower4pt\hbox{\hskip1pt$\sim$}}
    \raise1pt\hbox{$<$}}}         
\def\gsim{\mathrel{\rlap{\lower4pt\hbox{\hskip1pt$\sim$}}
    \raise1pt\hbox{$>$}}}         

\textwidth              16.4cm
\oddsidemargin           2.5cm
  \advance\oddsidemargin  by -1in
\evensidemargin          0.0cm
  \advance\evensidemargin by -1in
\marginparwidth          1.9cm
\marginparsep            0.4cm
\marginparpush           0.4cm
\topmargin              -0.5cm
  \advance\topmargin      by -0.5in
\textheight             24.0cm

\title{Do the E866 Drell Yan data change our picture of the chiral
structure of the nucleon ?}

\author{ N.N.~Nikolaev$^{a,b}$, W.~Sch{\"a}fer$^{a}$, A.~Szczurek$^{c}$, 
and J.~Speth$^{a}$ }

\address{ $^{a}$IKP(Theorie), FZ J{\"u}lich, D-52425 J{\"u}lich, Germany. \\
$^{b}$L.D. Landau Institute for Theoretical Physics, GSP-1, 117940,
ul. Kosygina 2, Moscow V-334, Russia. \\
$^{c}$ Institute of Nuclear Physics, PL--31--342 Cracow, Poland.}

\maketitle

\begin{abstract}
We revisit the evaluation of the pionic mechanism of the 
$\bar{u}-\bar{d}$-asymmetry in the proton structure function.
Our analysis is based on the extended AGK unitarity relation 
between contributions of different mechanisms
to the inclusive particle production and the total
photoabsorption cross-section (i.e. the proton structure function).
We reanalyze the role of isovector reggeons in inclusive production of
nucleons and Delta isobars in hadronic reactions. We find rather
large contribution of reggeon-exchange induced production of Delta isobars. 
This leaves much less room for the pion-exchange induced mechanism of
$\Delta$ production and provides a constraint on the $\pi N \Delta$ form factor.
The production of leading pions in proton-proton collisions at ISR
puts additional constraints on the $\pi NN$ vertex form factors.
An extension of the AGK-rules to reggeon exchange suggests
a negligible contribution to the proton structure function from
DIS off the exchanged $\rho,a_2$ reggeons.
All these constraints are used then to estimate the pion content
of the nucleon and allow to calculate parameter-free the $x$-dependence
of $\bar d - \bar u$.
We discuss the violation of the Gottfried Sum Rule and $\bar d$-$\bar u$
asymmetry and compare to the one obtained from the E866 experiment
at Fermilab. We estimate the background
to the pion structure function being determined by the ZEUS and H1
collaborations at HERA from leading neutron experiments.
\end{abstract}

\section{Introduction}

The violation of the Gottfried Sum Rule discovered by the NMC collaboration
at CERN in muon deep inelastic scattering \cite{A91} opened the long ongoing
discussion on the $\bar d - \bar u$ asymmetry in the nucleon. 
Extending and following the early work of Sullivan \cite{Sul72}, the asymmetry
can be naturally explained within the framework of the isovector 
meson cloud model of the nucleon (for recent reviews see \cite{ST98,Kum98} and
references therein), which uniquely predicted the asymmetry to be placed at large $x$,
in good agreement with the early NA51 Drell-Yan experiment \cite{NA51}. 
In the practical evaluation the dominant contribution can
be interpreted as due to the admixture of the $\pi N$ and
$\pi \Delta$ Fock-states in the physical proton.
The model results for the asymmetry depend in essential
way on the choice of the parameters
for $\pi NN$ and $\pi N\Delta$ vertices, and on the
elementary particle vs. Regge treatment of meson exchanges.
In most of the calculations the vertex parameters, which at present cannot
be calculated from first principles, were simply
adjusted to reproduce the observed Gottfried Sum Rule violation.
In contrast in Refs.\cite{Z92,HSS96} a unified approach to
hadronic reactions and deep inelastic scattering has been
pursued and the vertex parameters were constrained by the
experimental data on the leading nucleon and/or
leading Delta isobar production in high-energy hadronic reactions 
to which pion exchange contributes substantially.
The numerical evaluations of the $\bar{u}-\bar{d}$--
asymmetry were based on the fact that for an elementary particle exchange 
the contribution to the total
virtual photo--absorption cross section equals that to the inclusive
cross section for leading baryon production, which is one
of the manifestations of the extended AGK unitarity 
relations \cite{AGK74}. For the
pion exchange such a treatment is well justified,
because the reggeization effects are negligible. 

Besides the pion vertex parameters, the $\bar{u}-\bar{d}$--
asymmetry is affected by the contribution of heavier
isovector mesons.
In Ref.\cite{HSS96} the effect of heavier vector mesons
has been evaluated in an extended Fock state decomposition of the 
nucleon light-cone wave function into meson-baryon, including
$N\rho , N\omega , \Delta\rho$ in addition
to $\pi N , \pi \Delta$ and $\eta N$ which are
known to be important at low and intermediate
energy hadron scattering.

A fit to leading baryon experimental data suggests
then a rather large $\rho N$ component in the expansion of the nucleon
wave function. The relatively heavy $\rho$--meson,
carries then a large fraction of the nucleon momentum
and with a plausible ansatz for the $\rho NN$ vertex,
it introduces the $\bar d - \bar u$
which extends to rather large $x \gsim 0.5$ \cite{HSS96}.
The so-constructed model was perfectly consistent with the NA51 experiment at CERN
\cite{NA51,SEHS96}, which measured the
$\bar{u}-\bar{d}$--asymmetry at $x=0.18$ \cite{NA51}. Furthermore, after
the incorporation of the NA51 result, the global parton model analyses
have produced parametrizations which were practically indistinguishable
from the results of the meson cloud dynamical calculations
in \cite{HSS96}.  

A recent E866 experiment at Fermilab has reported a first high
precision mapping of the $x$-dependence of the $\bar{u}-\bar{d}$
asymmetry from the comparison of the $pp$ and $pd$ Drell-Yan 
production with the striking finding that the difference of the
$\bar{u}$ and $\bar{d}$ distributions seems to vanish at large $x \gsim 0.3$
\footnote{The determination of the $\bar{u}-\bar{d}$ difference 
in \cite{E866} depends on assumptions on the flavour symmetric sea 
$\bar{u} + \bar{d}$. Within the error bars of E866 one can regard 
$\bar{u} + \bar{d}$ as well constrained by the neutrino experiments,
the difference between $\bar{u} + \bar{d}$ from different 
gobal parton analyses is marginal.},
and is definitely smaller than the results of the model \cite{HSS96}
and of all the early global parton parametrizations based
on the NA51 result and the observed GSR violation.
There is no consensus yet on understanding this interesting result
\cite{Peng,MST98}. In the framework of the meson cloud picture,
one possible improvement is a Regge treatment of heavy particle exchanges,
which seems to be more appropriate in the kinematical region relevant
to the  $\bar{u}-\bar{d}$ asymmetry problem.
Although, because of the lack of a microscopic
QCD picture of reggeon exchange one is forced to use
rather a phenomenological approach, one can take advantage 
of a large body of work on the Regge phenomenology 
of hadronic two-body exclusive \cite{KS76,IW77}, and inclusive \cite{K79}, reactions,
which constrains the Regge vertices, and imposes useful constraints 
between Reggeon contributions to inclusive production of different baryons.
 
Recently there was some interest in understanding the role of
isoscalar reggeons in diffractive lepton deep-inelastic scattering
\cite{NSZ96,GK97} and in inclusive production of leading protons
at HERA \cite{SNS98}.
The role of isovector reggeons was discussed both in the context
of leading neutron production in electron DIS \cite{KPP96} and
in the context of diffractive DIS scattering with leading neutrons
\cite{GKS97}. As a matter of fact, the isovector Reggeon exchanges in
neutron and proton production must be much more important than in the evaluations of 
\cite{KPP96,GKS97} based on the total cross section analyses, because
the latter constrain only the spin--non--flip Regge vertices, and the much
more dominant spin--flip contribution has not been considered in \cite{KPP96,GKS97}. 
As will be discussed in the present paper this has important consequences
for the inclusive spectra of neutrons from $p \to n$ transitions
both in hadronic and lepton DIS and indirectly also for $\bar d - \bar u$
asymmetry.

In the present paper we discuss the role of isovector reggeon
exchange in the production of leading nucleons and $\Delta$ isobars.
The subsequent decay of the $\Delta$ isobar into $\pi N$ channel generates
additional nucleons both in hadronic as well as in lepton DIS processes.
Based on the extended AGK unitarity relation,
we shall argue however that in spite of the important role
in populating the $nX$ and $\Delta X$ inclusive channels,
the opening of the $nX$ and $\Delta X$ intermediate states
via Reggeon exchange leads to a negligible contribution to the 
hadronic total
cross sections and/or the structure functions in DIS.
As a consequence
the isovector reggeon contribution to the Gottfried Sum Rule violation
is rather small. On the other hand the reggeon exchange contributions
to the leading baryon spectra leave less room
for the pion exchange contributions, which are essential for
a quantitative understanding of
the Gottfried Sum Rule violation and the $\bar u - \bar d$ asymmetry.

Our principal conclusion is that once the wealth of information
on leading pions and leading neutrons and $\Delta$'s is accounted for
there emerges a consistent description of the E866 (NuSea) data.

\section{Inclusive production of pions, nucleons and Delta isobars
         in hadronic reactions}

\subsection{Recalling the Regge phenomenology}

Following Sullivan \cite{Sul72}, we expect that the dominant 
mechanism of leading neutron production in DIS is an absorption of the
virtual photon on the pion from the $\pi^+ n$-Fock state
of the nucleon, in which the
spectator neutrons are observed as leading neutrons. 
But the {\it{same}} dynamics is supposed
to be at work in the production of leading neutrons in hadronic
semi-inclusive reactions, where the virtual photon 
has been swapped for the hadronic projectile.
In the meson-baryon Fock-state picture of the proton's light-cone
wave function one is still left with nonperturbative parameters,
the 'radii' of the Fock-states or in other words the form factor
cut-off parameters. These parameters cannot be obtained from first
principles, but one may hope to constrain them in a reasonably reliable
way by demanding a consistency with experimental data for
hadron production at high energies.
This strategy has been taken in the work of the J\"ulich group
\cite{HSS96} and the earlier work by Zoller \cite{Z92} to
extract the parameters of vertex form factors.
Some subtle points connected to the distortion of the waves
(or absorptive or screening corrections) have been discussed in a recent 
paper \cite{NSZ97}, we shall also comment on that issue below.

\begin{center}
\begin{figure}
\epsfig{file=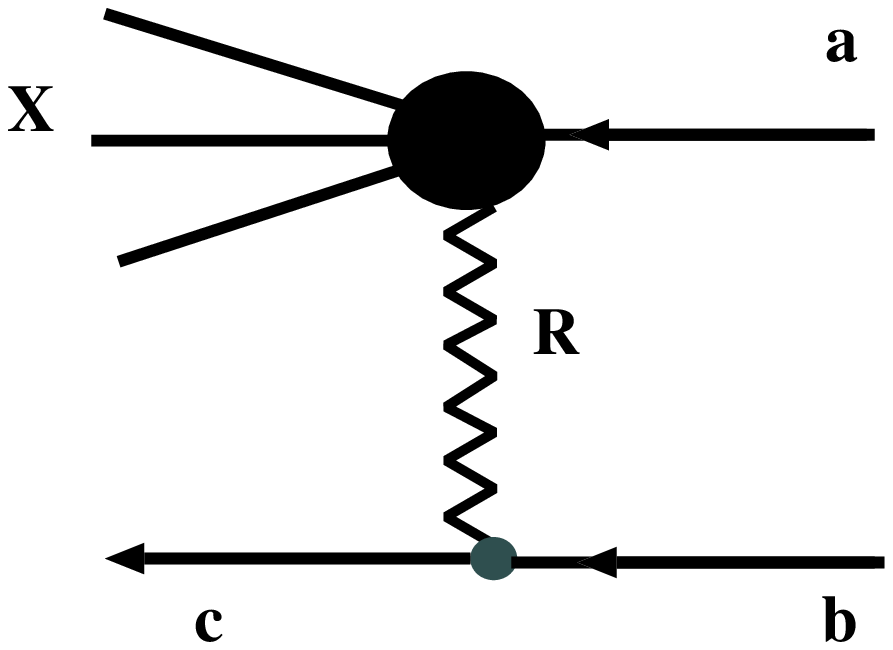, height = 6.0cm}
\caption{\it The inclusive production of particles $c$
in the reaction $ab \to cX$.}
\label{diagram1}
\end{figure}
\end{center}

Let us first collect the pertinent formulae for the following
discussions of the experimental data.
Following the standard phenomenology of inclusive
reactions, we define the Lorentz-invariant cross section 
(the so-called inclusive structure function) for the $a + b \rightarrow c + X$
reaction (see fig.\ref{diagram1})

\bea
\Phi^{b \to c} (z,\pt2) = E {d \sigma \over d^3\vec{p}}(ab \to c X)
= {z \over \pi}{d\sigma \over dz d\pt2}
= {1 \over \pi}{d\sigma \over dz dt} 
=  {s \over \pi}{d\sigma \over ds_X dt} 
\nonumber \\
= {1 \over 2 (2 \pi )^3} {1 \over 2 s} \sum_{i,X} \int d \tau_X  \;
\left | A^{ab \to cX}_i (s,t,s_X,\tau_X) \right |^2 \; .
\label{inclusive}
\eea
Here $(E,\vec{p})$ is the four-momentum of the outgoing particle $c$,
$z = p_z^{c.m.}/p_{max}$ is the longitudinal momentum fraction
(Feynman variable) of $c$ which for large
$z$ is identical to the lightcone variable,
 and $\vec{p}_{\perp}$ its transverse momentum, $t = -\pt2/z + t_{min}$ 
with $t_{min} = - (m_c^2-m_a^2)(1-z)/z - m_a^2 (1-z)^2/z$.
By $s_X$ we denote the invariant mass squared of the inclusive system $X$,
and $\tau_X$ is its Lorentz-invariant phase space.
Here the Mandelstam variables $s,t$ refer to the reaction $ab \to cX$;
and the index $i$ labels the exchange mechanism.
\footnote{Note that in Eq.(\ref{inclusive}) we have omitted the
contributions from
interferences of different exchange mechanisms $i,j$. They are absent in the
reactions considered as will be discussed below.}
In anticipation of large contributions from the region $1-z \ll 1$ to
the considered cross section we shall use the Regge form throughout
the present paper. The large Regge parameter in this case is
$s/s_X = 1/(1-z)$ and one can write the amplitudes $ A^{ab \to cX}_i $
in the form
\begin{equation}
 A^{ab \to cX}_i (s,t,s_X,\tau_X) = g_{ac}^i(t) \left( {s \over s_X}
\right )^{\alpha_i(t)} \eta(\alpha_i(t)) \cdot v^{ib \to X} (\tau_X,s_X) \; ,
\label{Regge-Amp}
\end{equation}
where $\alpha_i (t)$ is the Regge trajectory, $\eta_i (t) =
- [1 + \tau \exp(-i \pi \alpha_i (t))]/ \sin \pi \alpha_i (t)$ is the signature 
factor for the trajectory with signature $\tau$, and 
$ v^{ib \to X} (\tau_X,s_X)$ is the vertex for the transition
$ib \to X$ that leads to the (operational) definition
of the $i b$-- total cross section via
\begin{equation}
\sigma_{tot}^{ib} ={ 1 \over 2s_X} \, \sum_X
\int d \tau_X  |v^{ib \to X} (\tau_X,s_X) |^2 \, .
\end{equation}
With these definitions we finally obtain for the inclusive cross section
\begin{equation}
\Phi^{b \to c} (z,\pt2)  
= {1 \over 2(2\pi )^3} \sum_i 
\left( g^i_{ac}(t) \right)^2 \left| \eta (\alpha_i (t)) \right |^2
\left( {s \over s_X} \right ) ^{2\alpha_i(t) -1 } \, \sigma_{tot}^{ib} (s_X) \; .
\label{Regge_inclusive}
\end{equation}
At this point it is convenient to introduce the notion of a flux 
associated with the exchanged object in the $t-$channel:
\be
f_{i/c} (z_i) = \int d^2\vec{p}_\perp {E d \sigma^{(i)} (ab \to cX) \over \sigma^{ib}_{tot} d^3 \vec{p}} \;
\ee 
or more explicitly:
\be
f_{i/c} (z_i) = \int d\pt2 {\left( g^i_{ac}(t) \right)^2 \left| \eta (\alpha_i (t)) \right |^2 \over 16 \pi^2}\left( { 1\over z_i } \right)^{2\alpha_i(t) - 1} \; .
\ee
Here $z_i = 1-z = s_X/s$ is the longitudinal momentum fraction 
flowing through the $t-$channel. The function $f_{i/b} (z_i)$
has an interpretation of a flux of '$i$-quanta' emitted
by the hadron $a$ in the $a \to c$ transition. Correspondingly its integral
\be
n_{i/c} = \int_0^1 dz_i f_{i/c} (z_i)
\ee
shall be addressed as the number of '$i$--quanta' in $b$.

The contribution from the $t-$channel pion exchange Born-term to the
production of leading neutrons reads
\begin{equation}
\Phi_{\pi}^{p \to n} (z,\pt2)  = \frac{g_{pn\pi^+}^2}{16 \pi^3}
\frac{ (-t) }{(t-m_{\pi}^2)^2}
F^2_{\pi NN}(t)(1-z)^{1-2 \alpha_{\pi}(t)} \cdot \sigma_{tot}^{\pi a}(s_X). 
\label{pi_nuc}
\end{equation}
Here $\alpha_\pi (t)=  \alpha_{\pi}'(t-m_{\pi}^2)$ is the pionic Regge-trajectory,
$\alpha_\pi' \approx 0.7 \, $GeV$^{-2}$, $s_X = s(1-z)$ is the $a \pi$ cm-energy
squared, and  $g_{pn\pi^+}^2/4 \pi = 27.5$. Furthermore
$ t = -\pt2/z + t_{min}$ with $t_{min} = - (1-z)^2 m_p^2 /z$. $F_{\pi NN}(t)$
is the phenomenological vertex form factor that accounts for
the finite size of particles involved and/or off-shell effects.
Because of the proximity of the physical pion pole, the reggeization effects,
i.e. the departure of $\alpha_\pi (t)$ from $J_\pi = 0$ are
marginal and one can use $\sigma_{tot}^{\pi a} (s_X )$ for the on--mass
shell pion. 
The analogous formula for the $\Delta (1232)$ production reads:
\begin{equation}
\Phi_{\pi}^{p \to \Delta^{\tiny{++}}} (z,\pt2)  = \frac{g_{p\Delta^{++}\pi^-}^2}{16 \pi^3}
\frac{ (M_{+}^2-t)^2 (M_{-}^2-t) }
{6 m_{\Delta}^2 (t-m_{\pi}^2)^2}
F^2_{\pi N \Delta}(t)(1-z)^{1-2 \alpha_{\pi}(t)} \cdot \sigma_{tot}^{\pi a}(s_X) \; , 
\label{pi_del}
\end{equation}
where we introduced $M_{\pm} = m_{\Delta} \pm m_N$, 
$g_{p\Delta^{++}\pi^-}^2/4 \pi = 12.3 \, \mbox{GeV}^{-2}$,
and $ t = -\pt2/z + t_{min}$ with
$t_{min} = - (m_{\Delta}^2 - m_p^2)(1-z)/z - m_p^2 (1-z)^2/z$.
Again $F_{\pi N \Delta}(t)$ is the vertex form factor, a phenomenological quantity,
that has a priori no simple relation to its counterpart in Eq. (\ref{pi_nuc}).
Integrating Eqs.(\ref{pi_nuc},\ref{pi_del}) over $\pt2$ we get
\bea
z {d \sigma^{(\pi )} \over dz}(pp \to n X) = \pi \int d\pt2 \Phi_{\pi}^{p \to n} (z,\pt2)
  = {2\over 3} f_{\pi/N}(1-z) \cdot \sigma_{tot}^{\pi p}(s_X) , \nonumber \\
z {d \sigma^{(\pi )} \over dz}(pp \to \Delta^{++} X) = \pi \int d\pt2 \Phi_{\pi}^{p \to \Delta^{++}} (z,\pt2)
  = {1\over 2} f_{\pi / \Delta}(1-z) \cdot \sigma_{tot}^{\pi p}(s_X) \; .
\eea
%

\subsection{Inclusive spectra of pions}

For the moment let us focus on yet another issue, which did
not receive due attention in the literature. 
The Regge phenomenology outlined in section 2.1 allows to constrain the
large-$z$ behaviour of the vertex functions from the experimental
data on leading baryons. For smaller $z$ one must look into another constraint,
here the light-cone wave function formalism offers an 
intimate relationship between the flux of mesons originating
from a meson-baryon $MB$-Fock state and the corresponding
flux of baryons.
Clearly, if $f_{M/B}(z_M)$ denotes the probability to find
in the $MB$-Fock state a meson $M$ carrying the (light-cone)
momentum fraction $z_M$ and if $f_{B/M}(z_B)$ has the
analogous meaning of finding the baryon $B$, carrying
momentum $z_B$, it must hold true that 
\be
f_{M/B}(z) = f_{B/M}(1-z).
\label{symm}
\ee 
The probabilistic interpretation is obvious. 
The symmetry relation (\ref{symm}) allows us to write 
the contribution from spectator pions to the cross section 
for the production of leading {\it{pions}} as:
\begin{equation}
z_\pi {d \sigma^{(p)} \over dz_\pi}(pp \to \pi^0 X ) = 
{1\over 3} f_{N / \pi}(1-z_\pi) \cdot \sigma_{tot}^{pp}(s_X)=
 {1\over 3} f_{\pi/N}(z_\pi) \cdot \sigma_{tot}^{pp}(s_X) \; ,
\label{zdsigdz}
\end{equation}
where we focussed solely on the leading pions originating from the $\pi N$-Fock state,
a contribution from the $\pi \Delta$-states does not affect the following
conclusions.

\begin{center}
\begin{figure}
\epsfig{file=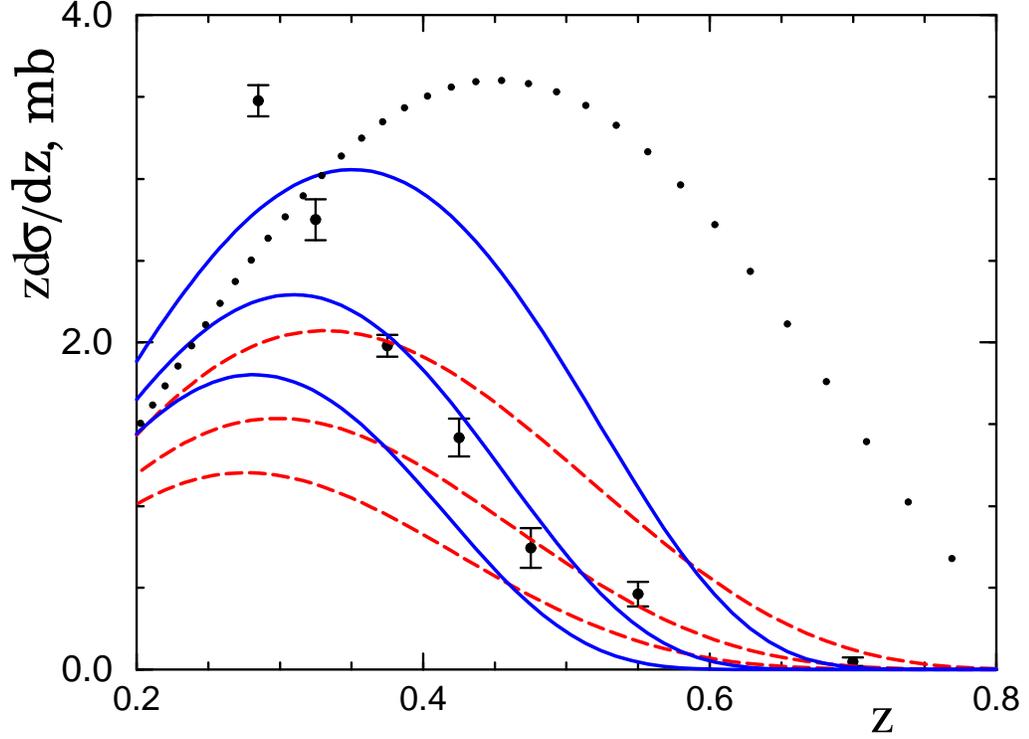, height = 10.0cm}
\caption{\it Differential cross section $zd\sigma /dz (pp \to \pi^0 X)$
at $p_{LAB} = 400 \, \mbox{GeV/c}$. The data are taken from \protect\cite{LEBC-EHS}.
The dotted curve shows a prediction of the model \protect\cite{HSS96}.
The solid curves were calculated with a 'Gaussian' form factor
$F_{\pi NN}(t) = \exp (-\left[R_G^2(t-m_\pi^2)\right]^2)$.
The curves in the figure correspond to $R_G^2 = 1.0\, ,1.5 , and \, 2.0 \; GeV^{-2}$
(from top to bottom). The dashed curves were calculated with an exponential
form factor $F_{\pi NN}(t) = \exp (R_E^2(t-m_\pi^2))$.
The curves in the figure are for $R_E^2 = 1.0\, ,1.5,and \, 2.0 \; GeV^{-2}$
(from top to bottom).}
\label{forward-pi}
\end{figure}
\end{center}

In the absence of a theoretical tool to directly calculate
the form factor  $F_{\pi NN}(t)$, 
the parametrization of $F_{\pi NN}(t)$ can only be judged from its
phenomenological success and consistency in a broad spectrum
of processes.
In the literature is
rather bold extrapolations from low energy physics have been habitual
 and leading pion production data have been seemingly overlooked in
previous considerations. As an example in figure \ref{forward-pi} 
we show the differential
cross section $z_\pi d\sigma / dz_\pi $ for production of the $\pi^0$-mesons
taken from the NA27 $pp$ experiment \cite{LEBC-EHS}. 
The spectra of $\pi^+$ and $\pi^-$ from the same experiment show
that the contribution from $\pi \Delta$ states is much smaller.
Shown are several theoretical curves,
calculated from Eq.(\ref{zdsigdz}) with different functional forms for the
$\pi NN$ form factor. The dotted curve shows the prediction from
the model of Holtmann and two of the authors \cite{HSS96}. In this
work a simple light-cone parametrization of the form factor had been adopted:
\begin{equation}
F_{\pi NN}(z,\pt2 ) = \exp \left[ -R_{LC}^2(M^2(z,\pt2) - m_N^2) \right] 
= \exp \left[ { R_{LC}^2(t-m_\pi^2) \over (1-z)} \right ] \; .
\end{equation}
\label{LCFF}
Here $M^2(z,\pt2) = (\pt2 + m_N^2)/z + (\pt2 + m_\pi^2)/(1-z)$ is the
invariant mass of the $\pi N$ Fock state, and
$R_{LC}^2 \approx 0.5 \mbox{GeV}^{-2}$ \cite{HSS96}. In the spirit of the
lightcone picture, the formfactor (\ref{LCFF}) was meant
to describe the low--mass components $M^2 \sim (m_p + m_\pi )^2$
and $1-z \sim m_\pi / m_p$. If one streches this simple
parametrization also to large $z_\pi = 1-z$ then one would
run into serious conflict with the experimental data on 
leading pions. As we shall see below, such an overprediction
of the leading pion spectra leads to a related overprediction
of the $\bar{u}-\bar{d}$ asymmetry at large $x$. 
The dipole (monopole) parametrization of the light-cone form factor
\cite{MST98} does not resolve the conflict with the NA27 data.
We have neglected here the absorptive corrections, but one would expect
them to reduce the Born term by at most a factor of $1.5-2$, see
for instance \cite{NSZ97}.

Clearly there is no real conflict between theory and experiment,
one simply should not extrapolate to large $z_\pi$ the 
functional form of form factors designed for
$z_\pi \ll 1$. Reversing the attitude, one should
look for parametrizations which are consistent with the leading
pion data and to explore the resulting constraints
from the leading pion data for predictions of the $\bar{u}-\bar{d}$
asymmetry in DIS. 
We show the result for two different simple options only,
the exponential parametrization $F_{\pi NN}(t) = \exp (R_E^2 (t-m_\pi^2))$
(dashed lines, for values of $R_E^2 = 1.0, 1.5, 2.0 \, \mbox{GeV} ^{-2}$)
and the 'Gaussian' parametrization
$F_{\pi NN}(t) = \exp (-\left[R_G^2(t-m_\pi^2)\right]^2)$ (solid lines,
for values of $R_G^2 = 1.0, 1.5, 2.0 \, \mbox{GeV} ^{-2}$).

We conclude that the forward pion data suggest that pions
in the $\pi N$ Fock state of the nucleon wave function are rather "soft",
carrying longitudinal momenta of not more than $z \approx 0.5$.
This is quite a subtle constraint on the tail of
the $\pi N$ two-body wave function and imposes
 the empirical bounds $R_{E,G}^2 > 1 \, \mbox{GeV}^{-2}$.


\subsection{Inclusive spectra of neutrons from $p \rightarrow n$}

\begin{center}
\begin{figure}
\epsfig{file=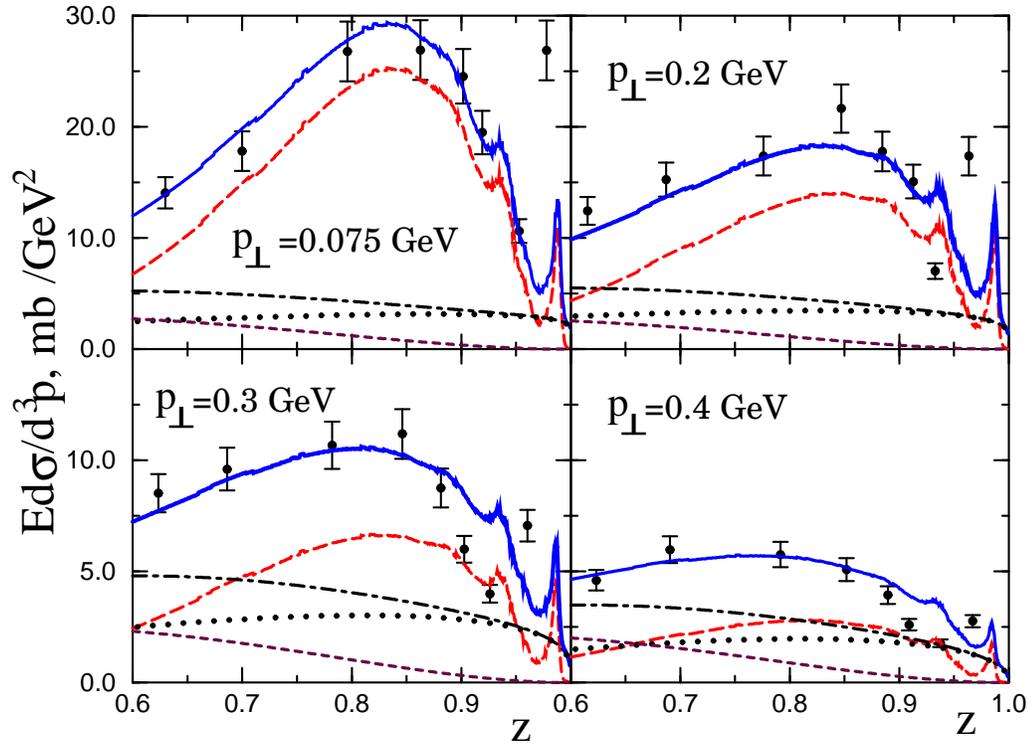, height = 10.0cm}
\caption{\it Invariant cross section for the reaction
$pp \to nX$ at $p_{LAB} = 24 \, \mbox{GeV/c}$.
The experimental data are taken from \protect\cite{blobel}.
The long dashed curve shows the contribution from the pion exchange;
the dotted curve is the $\rho,a_2$-exchange
contribution, and the dashed curve shows the contribution from the two
step process $p \to \Delta \to n$. In additionally we present the the sum of
the two background contributions as the dot-dashed line.
Finally the solid curve represents the sum of all components. }
\label{Blobel}
\end{figure}
\end{center}

\begin{center}
\begin{figure}
\epsfig{file=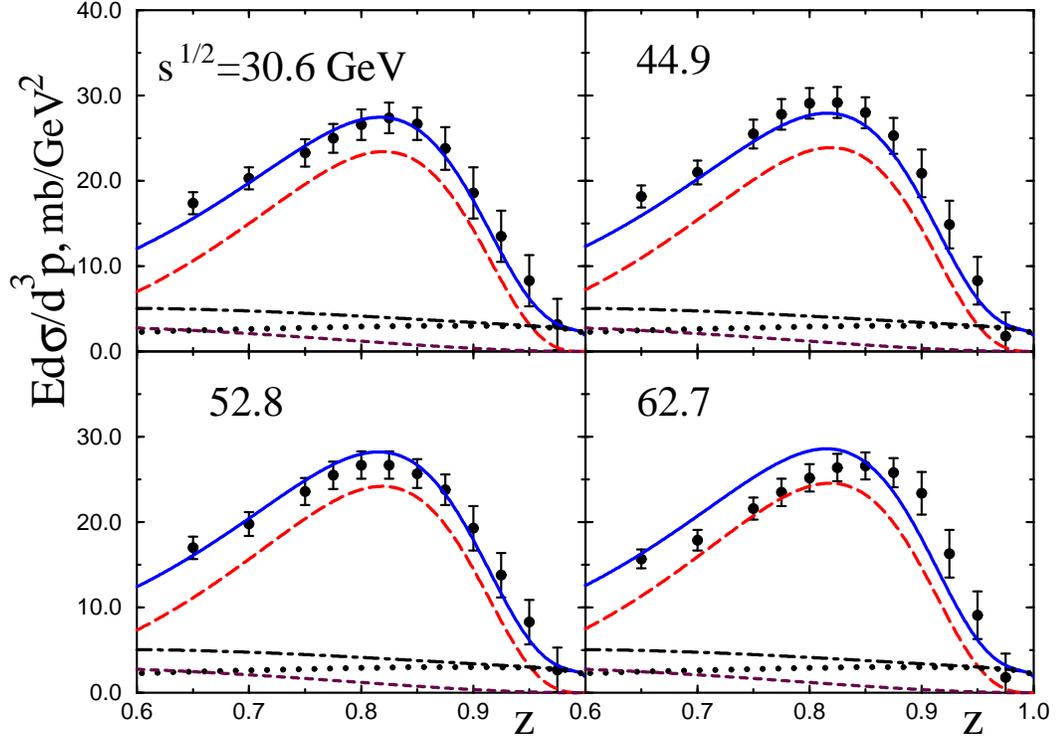, height = 10.0cm}
\caption{\it Invariant cross section for the reaction
$pp \to nX$ as a function of z for $\pt2 =0$.
The long dashed curve shows the contribution from the pion exchange;
the dotted curve is the $\rho,a_2$-exchange contribution,
and the dashed curve shows the contribution from the two-step process
$p \to \Delta \to n$. In addition we present the sum of
the two background contributions as the dot-dashed line.
The solid curve represents the sum of all components. The experimental
data are taken from \protect\cite{FM76}.}
\label{high-energy}
\end{figure}
\end{center}

Let us proceed now to the neutron spectra.
In what follows we shall use the 'Gaussian' parametrization with a
cut-off parameter $R^2_G = 1.5 \, \mbox{GeV}^{-2}$ which does a reasonable job
for the leading pion data (see Fig.(\ref{forward-pi})).
In Figs. (\ref{Blobel}, \ref{high-energy}) we show the relevant data
for the forward neutron production.
In the analysis of these data we also improved on what has been done before.
Absorptive corrections to the $t-$channel pion exchange are incorporated
in terms of standard methods of the Reggeon calculus in the
quasi-eikonal formulation. The formalism is presented in \cite{NSZ97}
and there is no need to repeat that here.

For the $\pi N$ total cross section $\sigma^{\pi N}_{tot}(s_x )$ 
we can use the convenient database provided in 
\cite{PDB96}. It is especially convenient at small $s_X$ where
the resonance structure of $\sigma^{\pi N}_{tot}(s_x )$ is
important, as this is the case with the low--energy 
data in \cite{blobel} ($\sqrt{s} = 4.93, 6.84 GeV$).

We decompose the inclusive structure function for neutron production
in terms of the pion and background contributions:
\begin{equation}
 \Phi^{p \to n} (z,\pt2) =  \Phi_{\pi}^{p \to n} (z,\pt2) +
\Phi^{p \to n}_{\rho , a_2}(z,\pt2) +
\Phi^{p \to n}_{\Delta}(z,\pt2) \; .
\label{incl}
\end{equation}
We take into account two background contributions:
exchange of the $\rho, a_2$-trajectories (which are assumed
exchange-degenerate following the Regge theory wisdom)
and production of neutrons
via the two-step processes $p \to \Delta \to n$.
The deacy of $\Delta$'s had been entirely 
neglected in all previous analyses except
in Ref.\cite{SNS98} where it was shown to play an important role for
leading proton production. 
%
%
The contribution of this mechanism to $p \to n$
reaction is suppressed in comparison to $p \to p$ reaction if the dominant
mechanism of the $\Delta$ production is an exchange by
an isovector object.
In the spirit of the Regge phenomenology the contribution of
$\rho,a_2$ to (\ref{incl}) is parametrized as
\begin{equation}
\Phi^{p \to n}_{\rho , a_2}(z,\pt2) = 
\sigma_R 
({\pt2 \over 4 m_p^2} + \chi^2)
(1-z)^{-2 \alpha_\rho' t}
\exp \left(-B_R {\pt2 \over z}\right).
\end{equation}
Please note that the $\rho - a_2$ interference is negligible because
of the different G-parity of $\rho$ and $a_2$.
The non spin-flip/spin-flip ratio is known to be small:
$\chi \approx 0.12$ \cite{AN84}. In accord with the Regge phenomenology
we take $\alpha'_R = 1$ GeV$^{-2}$.

There are no unique extrapolations from the $\rho,a_2$ particle pole to the 
Reggeon exchange and $\sigma_R$ remains a free parameter.
Our analysis of
the neutron data in the large $z$ region suggests
$\sigma_R \approx 250 \, \mbox{mb GeV}^{-2}$
(for the slope a typical value of $B_R \approx 6.5 \; \mbox{GeV}^{-2}$
was taken), which is similar to earlier estimates \cite{ZS83}.

In evaluation of the background coming
from the decay of fast $\Delta$ resonances,
one can either model the spectra of $\Delta$ isobars,
as done in Ref.\cite{SNS98}, or take any convenient parametrization, which 
reproduces the experimental data \cite{LEBC-EHS} .
The decay of $\Delta$'s lead to a significant shift in $z$ distributions,
compared to initial $\Delta$ distributions and only moderate modification
in perpendicular momentum distributions. We have found that the 
spectrum of neutrons from $\Delta$-decays can
be parametrized to a good approximation by the following simple form
\begin{equation}
\Phi^{p \to n}_{\Delta}(z,\pt2) = 
\sigma_0 z^3 (1-z)^3 \exp(-B_\Delta \pt2)
\end{equation}
with $\sigma_0$ = 48 mb GeV$^{-2}$, $B_\Delta = 2 \; \mbox{GeV}^{-2}$.

We find that there is substantial correlation between 
the parameter of the form factor and
the strength of the absorptive corrections. The choice of the form factor
parameter made above ($R_G^2 = 1.5 \, \mbox{GeV}^{-2}$) 
results in rather weak absorptive corrections
which has only a weak effect of a $10-20 \%$ reduction
of the pion-exchange Born-term, mildly depending on $z$ and $\pt2$.

The description of the experimental data is excellent in general.
The pion exchange is clearly seen to be the dominant mechanism
in the region of $z = 0.7 - 0.9$ and $\pt2 < 0.2-0.3 \, \mbox{GeV}^2$.
Because $\alpha_{\rho,a_2} (t) > \alpha_\pi (t)$, the Regge theory 
uniquely predicts the dominance of the Reggeon exchange 
at $z \to 1$ and the experimental data confirm that.
We improve upon an oversimplified treatment of the background 
in \cite{NSZ97}, where its contribution was modelled scaling
up the pion exchange contribution, which underestimates the $\rho ,a_2$
background at large z. In the earlier work of two of the authors 
\cite{HSS96}, the $\rho$ (particle-) 
exchange had been included within the light-cone formalism 
with the form factors parametrized in the form following 
eq. (\ref{LCFF}) which also underestimates the $\rho$ contribution 
at large $z$. 

Notice that especially in the region of large $z$ the
inclusion of $\rho,a_2$ Reggeons leads to much improved
description as compared to the previous calculations \cite{HSS96,NSZ97}
where only the pion exchange did contribute at large $z$. 
The fluctuations in the theoretical curves in Fig.\ref{Blobel} are due to
the use of experimental values of the cross sections.
Indeed the region of large z corresponds to rather small $s_X$ of
the inclusive system $X$, for example the peak at $z$ close
to unity corresponds to the $\Delta(1232)$--resonance in pion-nucleon scattering.

\subsection{Inclusive spectra of $\Delta^{++}$ from the
$p \rightarrow \Delta^{++}$ reaction}

By far less ideal is the situation with to the production
of forward $\Delta$'s. Experimental data is scarce, and if available often
of rather poor quality.
Many sets of data suffer from substantial ambiguities
from the nonresonant $\pi N$ background subtraction 
(a factor  $\approx 2$ uncertainty is typical, 
see for instance \cite{Higgins}) 
and cannot be used in the analysis.
We do not have at hand measurements of double differential cross
sections, that would allow us a meaningful determination of the 
$\pt2$-dependence of the individual mechanisms. 
In particular the data is not sufficient
to make any quantitative statement on the role of absorptive corrections.
Nonetheless the experimental data provide
valuable bounds on the $\pi N \Delta$ form factor which
translate into substantial constraints on the $\pi \Delta$ 
contribution to the $\bar d - \bar u$ asymmetry.

\begin{center}
\begin{figure}
\epsfig{file=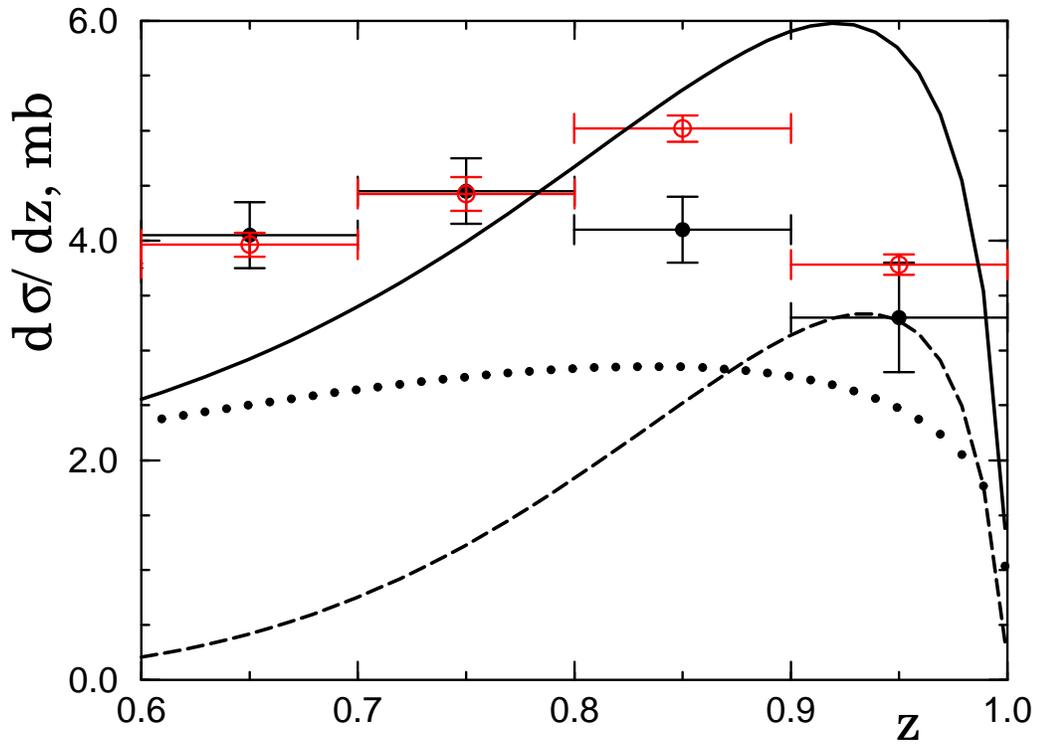, height = 10.0cm}
\caption{\it Differential cross section
$d\sigma / dz$ for the reaction $pp \to \Delta^{++} X$
at $p_{LAB} = 400 \, \mbox{GeV/c}$.
The dashed curve is the contribution from pion-exchange; the
dotted curve shows the $\rho ,a_2$ contribution.
Shown by the solid curve is the sum of the two.
Experimental data are taken from \protect\cite{LEBC-EHS}(open circles),
and \protect\cite{ABCDHW} (filled circles).}
\label{Delta}
\end{figure}
\end{center}

In Fig.\ref{Delta} we show the $\pt2$-integrated cross section
$d \sigma / dz (pp \to \Delta^{++} X)$
from two CERN experiments obtained by the ABCDHW \cite{ABCDHW} and
LEBC-EHS \cite{LEBC-EHS} collaborations.
The upper bound to the pion exchange can be obtained by adjusting the
parameter of the form factor in Eq.(\ref{pi_del}) to the CERN data
\cite{ABCDHW,LEBC-EHS}.
Based on the Regge factorization, we evaluate the 
background from the $\rho$ and $a_2$ reggeon exchanges as follows.
The experimental data on the two-body reaction $\pi^+ P \to \pi^0 \Delta^{++}$
exhibit a deep minimum of the differential cross section at $t=0$,
which is consistent with a pure $M1$-transition \cite{StS63,ZS78}.
The related data on the $\pi^+ p \to \eta^0 \Delta^{++}$ charge exchange
reaction show a similar dominance of the M1 transition in the $a_2 N\Delta$ vertex.

Therefore we take the following parametrization for the $\rho,a_2$ contribution:  
\begin{equation}
\Phi^{p \to \Delta^{++}}_{\rho , a_2}(z,\pt2) = 
\lambda_0 \sigma_R {\pt2 \over 4 m_p^2}
(1-z)^{-2 \alpha_\rho' t}
\exp \left(-B_R {\pt2 \over z}\right) \; .
\end{equation}
A careful inspection of two-body charge-exchange reactions
$\pi^- p \to \pi^0 n$ and $\pi^+ p \to \pi^0 \Delta^{++}$ in a broad
range of energy suggests (assuming Regge factorization)
that the $\rho p \Delta^{++}$-coupling strength
should be about a factor $1.5$ larger than that for the $\rho p n $ case.
Correspondingly, $\lambda_0 = 1.5 $. 
Consequently, the Regge phenomenology offers a parameter--free evaluation
of the $\rho,a_2$ exchange in production of leading $\Delta$'s in
terms of the $\rho,a_2$ exchange background to the production
of leading neutrons. To the best of our knowledge, this relationship has
not been used before.
Consistently with
the two-body reactions we take here the same slope parameter $B_R$
as for the neutron production. 
The contribution of the reggeon exchange is shown in Fig.\ref{Delta}
as the dotted line. As can be seen from the figure it exhausts
a significant fraction of the spectra leaving less room for the pion exchange
contribution. 

Let us return to the estimate of the pion exchange contribution
which we take in the form according to Eq.(\ref{pi_del}) with an 
exponential form factor
$F_{\pi N \Delta}(t) = \exp \left( R_\Delta^2 (t - m_\pi^2) \right)$.
Unfortunately the quality of the data \cite{ABCDHW,LEBC-EHS} does not
allow for a fit of $R_{\Delta}$.
 The difficulties with the inclusive $\Delta$ spectrum
have already been observed in the early work on the topic \cite{G74}, 
and without the constraints from the two-body reactions, one
might be tempted to neglect the $\rho$--contribution at all, we quote 
from Gotsman\cite{G74}:"In an attempt to {\it {improve the fits}} we added the exchange
of a $\rho$ trajectory, {\it {but to no avail}} [...]".
 Only a lower limit on $R_{\Delta}$ can be obtained by comparing
the sum of the reggeon exchange and pion exchange contributions
\begin{equation}
{d \sigma \over dz}(pp \to \Delta^{++} X) = {\pi \over z} \int d \pt2 
\left ( \Phi^{p \to \Delta^{++}}_{\pi}(z,\pt2)
+\Phi^{p \to \Delta^{++}}_{\rho , a_2}(z,\pt2) \right ) 
\end{equation}
to the experimental data \cite{ABCDHW,LEBC-EHS}.
We consider $R_\Delta^2 = 2 \; \mbox{GeV}^{-2}$ to be the lower limit on
the parameter. The corresponding upper limit of the pion exchange
contribution is shown by the dashed line in Fig.\ref{Delta}.
The pion exchange contribution found here is even smaller than
in Ref.\cite{HSS96} and excludes the scenario with large $\pi \Delta$
component discussed in \cite{MST98} as a possible explanation
of the restoration of $\bar u - \bar d$ symmetry at intermediate Bjorken-$x$
observed in \cite{E866}. This will have important consequences
for the Gottfried Sum Rule violation and $\bar d - \bar u$ asymmetry.

A digression into unitarity relations and/or the extended AGK rules 
is in order before we proceed with implications of the 
above analyses of inclusive reactions for the partonic structure
of protons as seen in DIS.

\section{From inclusive hadronic cross section to the total
cross section: the extended AGK rules}

As has been demonstrated in the previous section the reggeon exchanges
may contribute significantly to individual inclusive channels
and definitely influence the production of leading baryons.
The impact of specific inclusive channels on the total cross section
(total photoabsorption cross section for DIS) is controlled by unitarity.
The well known example is the so--called AGK rules for diffractive
scattering (Pomeron exchange)\cite{AGK74}, by which the opening of diffractive
channels gives rise to the shadowing correction and reduces the 
proton structure function \cite{Barone93,NZ94,BR97}. The gross features
of this relationship can be understood as follows.
In order to isolate how opening of the individual inclusive channels
$ab \to cX$ via specific exchange mechanisms
affects the total $ab$ cross section, let us consider 
the discontinuity of the $ab$--forward scattering amplitude associated
with the $c+X$ intermediate state (see fig.\ref{diagram1}).
The optical theorem relates its contribution to the total cross section
\begin{equation}
\Delta^{(2)}_i \sigma_{tot}^{ab} =
{1 \over s}\,  Im T^{(2)}_i (s,t=0)
\; .
\end{equation}
For the  $i$--type--Reggeon exchange mechanism of $c+X$ production
it can be calculated as
\footnote{see for instance ref.\cite{K79}.}
\bea
T^{(2)}_i(s,t=0) = {1 \over 2!} \int {d^4 k \over (2\pi )^4 i}\; {\left[ g^i_{ac} (k^2) \right]^2 \over (p-k)^2 - m_c^2 + i\epsilon} \;  \eta_i(k^2)^2 \left( {s \over s_X} \right) ^{2\alpha_i (k^2)} 
{\cal{T}} (ib \to ib) \, .
\eea 
Where the Reggeon-particle scattering amplitude ${\cal{T}} (ib \to ib)$ is
defined such that it fulfills the generalized unitarity condition
\be
Im \, {\cal{T}} (ib \to ib) = {1 \over 2} \int d\tau_X  \left| v^{ib \to X} (\tau_X,s_X)  \right|^2 
= s_X \sigma_{tot}^{ib} \, .
\ee
Utilizing this relation after integrating over the propagator pole of particle $c$ and a suitable change of integration variables, we obtain
\bea
T^{(2)}_i(s,t=0) = i \, \int {d^2 k_\perp ds_X \over 2 \, (2\pi )^3}\, \left( g^i_{ac} (k^2) \right)^2 \,  \eta_i (k^2)^2 \left ( {s \over s_X} \right)^{2\alpha_i (k^2) - 1}  \; \sigma_{tot}^{ib}(s_X )\, .
\eea 
Now we observe that $\eta_i(k^2)^2 = \tau \exp (-i \pi \alpha_i(k^2)) | \eta_i(k^2) |^2 \,$
and approximate the phase factor by its value at $k^2 = 0$. We can now easily establish (cf. eq. (\ref{Regge_inclusive})) the connection between the double
scattering contribution to the total cross section $\Delta^{(2)}_i \sigma \equiv Im T^{(2)}_i / s$ and the inclusive cross section:
\be
\Delta^{(2)}_i \sigma^{ab}_{tot} = Im \left( i \,\mbox{e}^{- 
 i \pi\alpha_i(0)} \right) \int {d^2 k_\perp ds_X \over \pi}  \, {d\sigma_i(ab \to cX)  \over ds_X dt} \equiv \xi_i \, \sigma^{incl}_i \, .
\label{extAGK}
\ee
where we defined
\be 
 \sigma^{incl}_i =  \int {d^2 k_\perp ds_X \over \pi}  \, {d\sigma_i(ab \to cX)  \over ds_X dt} \approx   \int dt ds_X  \, {d\sigma_i(ab \to cX)  \over ds_X dt}
\ee
Let us illustrate the implications of the extended AGK-rule
(\ref{extAGK}) by considering some examples:
\begin{itemize}
\item{for the pomeron, $\alpha_\Pom (0) \approx 1,\; \tau = +1$  
and thus $\xi_\Pom = -1$. This
is a well established fact, which has become known as one
of the AGK cutting rules, meaning that {\it{opening of diffractive channels}} 
leads to a {\it{reduction}} of, or the
absorptive correction to, the total cross section.}
\item{for the pion,  $\alpha_\pi (0) \approx 0,\; \tau = +1$, 
hence $\xi_\pi = +1$. This 
implies that inelastic interaction with pions in the target hadron 
$a$ enhances the total cross section by an amount $n_{\pi /a} 
\sigma_{tot}^{\pi b}$.}
\item{for the Reggeon with $\alpha_R (0) \approx 0.5$ ($\rho,a_2,f_0,\omega$),
$\xi_R = 0$; which means that the contribution from inelastic
interaction with Reggeons 'in the hadron $a$' to the total cross section
vanishes.}
\end{itemize}
The latter two results concerning the contributions from the $\pi$-- trajectory and the secondary Reggeons are also known in a different context for some time:
in calculations of nuclear shadowing in hadron-deuteron collisions it was observed that the
inelastic intermediate states excited by $\pi$-exchange contribute 
to an {\it{anti-}}shadowing, while the contributions 
from excitation of intermediate states by other
secondary Regge trajectories are negligible \cite{KK73,Dakhno83}.
  
\section{Pion content of the nucleon and $\bar d - \bar u$ asymmetry}

Having set the parameters of the pion flux factors
in section 2 we can proceed to the calculation of the corresponding quark
distributions. Those can be calculated as a simple convolution of the flux
factor and quark distributions in the pion
\begin{equation}
\Delta^{(\pi )}q_f(x,Q^2) = \int_0^x \frac{dz_{\pi}}{z_{\pi}}
f_{\pi}(z_{\pi}) \cdot q^\pi_f({x \over z_{\pi}},Q^2)  \; ,
\label{convolution}
\end{equation}
where $f_{\pi} = f_{\pi/ N}, f_{\pi / \Delta}$ as introduced in section 2.
In our calculation we take the GRV-pion structure function \cite{GRV92}
at the average value of the E866 experiment $Q^2 \approx$ 50 GeV$^2$. 
We focus on the difference
$\bar d(x,Q^2) - \bar u(x,Q^2)$. Due to flavour symmetries only valence
quark distributions in the pion contribute to this quantity. The less known sea
contribution cancels in $\bar d(x,Q^2) - \bar u(x,Q^2)$\cite{SHS96}.
While the quark distributions in the pion can be verified in the valence
region (intermediate and large $x$), the Drell-Yan processes do not allow
to determine them in the sea region (low $x$).

As discussed in the previous section the contribution of DIS off
exchanged $\rho ,a_2$--Reggeons
to the proton structure function is negligible
and will be omitted in the following.

\begin{center}
\begin{figure}
\epsfig{file=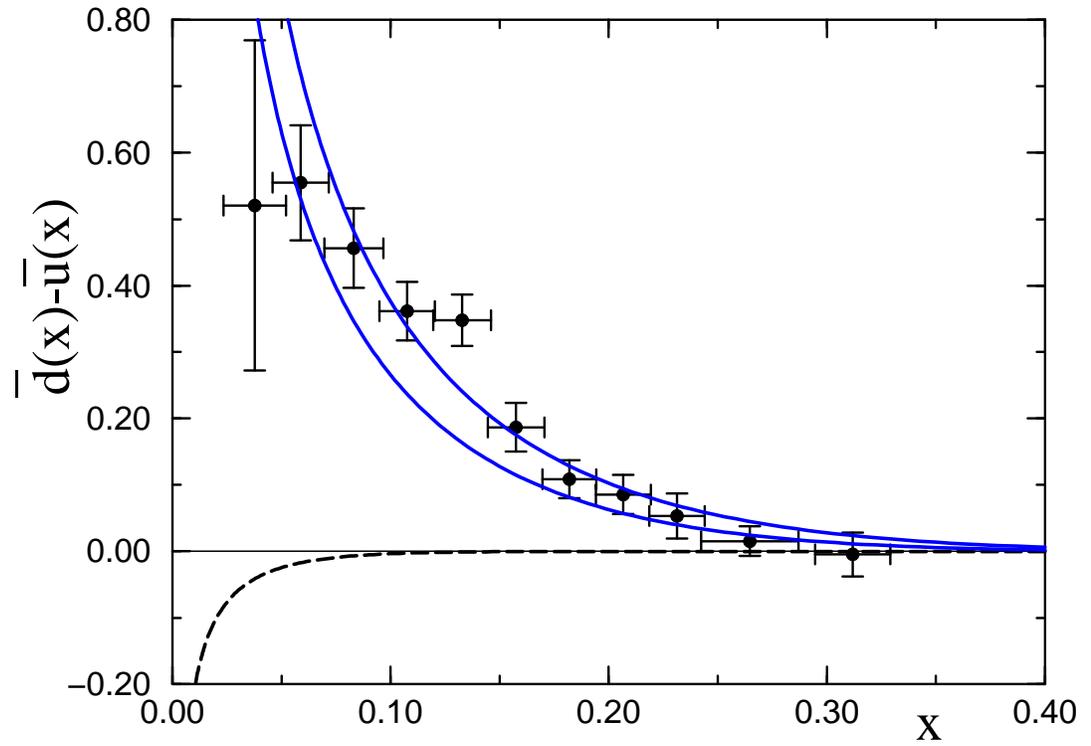, height = 10.0cm}
\caption{\it Flavour asymmetry $\bar{d} (x) - \bar{u} (x)$ at $Q^2 = 54 \, \mbox{GeV}^2$. Experimental data are from E866 \protect\cite{E866}. The solid curves show the contribution from the $\pi N$-Fock state and were calculated for Gaussian form-factors; the upper curve belongs
to $R_G^2 = 1 \, \mbox{GeV}^{-2}$, the lower one to  $R_G^2 = 1.5 \, \mbox{GeV}^{-2}$
The dashed line shows the contribution of the $\pi \Delta$-Fock state,
calculated for an exponential form factor with $R_{\Delta}^2 = 2 \;\mbox{GeV}^{-2}$.}
\label{dbar-ubar}
\end{figure}
\end{center}

In Fig.\ref{dbar-ubar} we display $\bar d (x) - \bar u (x)$
which is due to pionic contributions as obtained in the present analysis
 and compare it to the recent result of the NuSea
collaboration \cite{E866}. 
The $\pi \Delta$ contribution (dashed line) becomes important only at
rather small values of Bjorken-$x$ which is due to rather soft form factor
as suggested by the analysis of leading $\Delta$ isobars.
The analysis above clearly demonstrates that it is possible to construct
the pion flux factor consistent with both hadronic and Drell-Yan data,
provided the background processes in hadronic reactions are taken
carefully into account.

Let us finally comment on the violation of the Gottfried sum rule.
We obtain a number of pions in the proton
in the $\pi N$ Fock state of $n_{\pi N} = 0.21$ for
the cutoff parameter $R_G^2 = 1.5$ GeV${^{-2}}$ and
$n_{\pi N} = 0.28$ for  $R_G^2 = 1$ GeV${^{-2}}$.
For the $\pi \Delta$ Fock state, a cutoff $R_\Delta^2 =2$GeV$^{{-2}}$
yields  $n_{\pi \Delta}$ = 0.03.
The latter value is considerably smaller (by a factor 2 - 8) than
those obtained in all previous analysis.
These pionic multiplicities translate into the integrated value 
of $\bar d - \bar u$ asymmetry by means of
\begin{equation}
I_A = \int_0^1 [ \bar d (x) - \bar u (x)] \; dx =
\frac{2}{3} n_{\pi N} - \frac{1}{3} n_{\pi \Delta},
\end{equation}
yielding values of $I_A = 0.129$ for  $R_G^2 = 1.5$ GeV${^{-2}}$ and
 $I_A = 0.177$ for  $R_G^2 = 1$ GeV${^{-2}}$.
 
For the Gottfried integral
\begin{equation}
S_G = \frac{1}{3} - \frac{2}{3} \int_0^1  [ \bar d (x) - \bar u (x) ] \; dx
\end{equation}
we obtain correspondingly $S_G = 0.247$ for $R_G^2 = 1.5$ GeV${^{-2}}$
and $S_G = 0.215$ for $R_G^2 = 1$ GeV${^{-2}}$.
The value reported by NMC is $S_G = 0.235 \pm 0.026$ \cite{A91}
which is perfectly consistent with our result
$ 0.21 \lsim S_G \lsim 0.25 $.

\section{Non-pionic background in the production of leading
neutrons at HERA}

\begin{center}
\begin{figure}
\epsfig{file=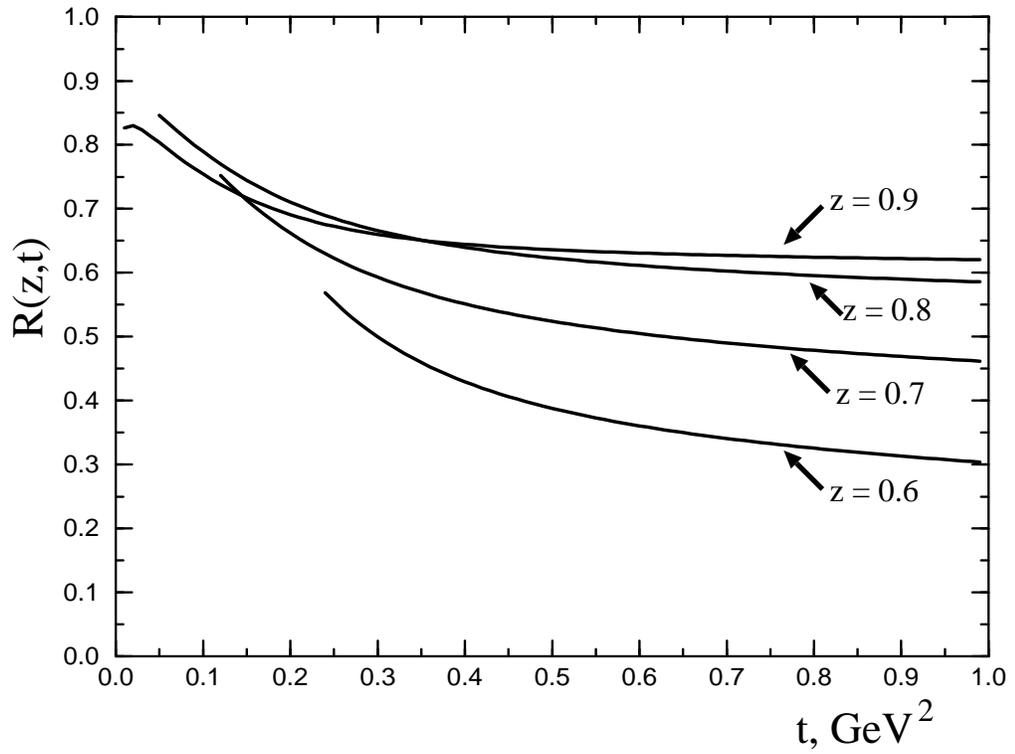, height = 10.0cm}
\caption{\it The signal to background ratio $R(z,t)$ as defined in
Eq.(\ref{signal-to-background})
shown as a function of $t$ for several values of $z$. Note that the
pion contribution peaks between $z=0.8-0.9$.}
\label{background}
\end{figure}
\end{center}

There is still another interesting spin off of the present analysis.
As discussed in section 2 the reggeon exchange contribution and decays
of the $\Delta$ resonances constitute a sizeable background
to the pionic contribution especially at large transverse
momenta of leading nucleons. The same peripheral
processes are at work in the production of leading neutrons in electron
deep inelastic scattering at HERA,where leading neutron tagged DIS
is presently studied experimentally as the method of determination
of the pion structure function \cite{HLNSS94,H1_neutron,ZEUS_neutron}.
The quality of the neutron tagging can be judged from the pion
purity factor shown in fig.\ref{background},
here we present the ratio $R(z,t)$ defined as
\begin{equation}
R(z,t) = {I^\pi (z,t) \over [ I^\pi (z,t) + I^{bg} (z,t)]} \, ,
\label{signal-to-background}
\end{equation}
where 
\begin{equation}
I^{\pi,bg}(z,t) = \int_{t_{min}}^t d\tilde{t} \,  \Phi^{p \to n}_{\pi,bg}(z,\pt2 ) \, ,
\end{equation}
and the index $bg$ denotes the 
background (reggeon exchange + $\Delta$ decays in our case).
The calculations for fig. \ref{background} were performed for 
the hadronic reaction $pp \to nX$. they can, however, also
be used as an estimate for the deep inelastic reaction 
$\gamma^* p \to nX$. One should be aware of the fact, that this amounts to an 
implict assumption of
\begin{equation}
\frac{F_2^R(x)} {F_2^{\pi}(x)} =
\frac{\sigma_{tot}^{Rh}} {\sigma_{tot}^{\pi h}} \;  .
\end{equation}
Certainly this relation is approximate and can be used as a guiding
principle only.
A better approximation is however beyond the present, 
rather phenomenological,
understanding of reggeon exchanges.

Our phenomenological analysis suggests that in order to extract the pion
structure function one should limit rather to low $t$ and/or low transverse
momenta of neutrons. This is the region where the leading proton spectrometers
of both H1 and ZEUS collaborations have the largest sensitivity.
In order to identify the "unwanted" background one could make the analysis
of the experimental data for different cuts in transverse momentum.
 
\section{Conclusion}

The NuSea E866 experiment provided the first detailed mapping of
the $x$-dependence of the $\bar{d}-\bar{u}$ asymmetry in the proton
sea. While confirming the global features of the asymmetry predicted
by the meson cloud picture, the large--$x$ behaviour of these data 
called for revisiting the model adding more constraints from 
applications of the same model to hadronic inclusive reactions. 
Specifically, in this paper we demonstrated that the experimental
data on fragmentation of protons into leading pions impose useful
constraints onto the pion iduced $\bar{d}-\bar{u}$ asymmetry such
that the asymmetry must be negligibly small at $x\gsim 0.3$ in
perfect consistency with the NuSea findings. 

Our reanalysis has been based on a unified treatment of inclusive
production of leading nucleons and $\Delta$'s in hadronic collisions
and we paid special attention to the background to pion exchange
from isovector $\rho,a_{2}$ exchanges. Then, based on the extended
AGK unitarity rules, we related the contributions of different
exchange mechanisms to inclusive $ab \to  cX $ cross section
to the contribution of inelastic interaction (DIS) on the
exchanged objects to the total $ab$ cross section (DIS structure function
of the target hadron). We observe that the extended AGK rules suggest
a negligible contribution of DIS off the exchanged $\rho , a_2$ 
to the $\bar{d}-\bar{u}$ asymmetry, leaving pions as the dominant
source of the asymmetry.
In numerical evaluation of the pion contribution to inclusive
cross section, and eventually to the asymmetry,
we made an extensive use of the Regge
factorization which allows to relate the $\rho,a_2$ exchange
contribution to the leading neutron and $\Delta$ production 
based on the Regge phenomenology of two-body charge exchange reactions.
One of the results is that the isovector $\rho,a_2$
reggeon exchanges exhaust a large fraction of inclusive leading 
$\Delta$ production. The implication for the asymmetry is a
substantial reduction of 
the contribution from the $\pi \Delta$ Fock states which. As a 
result a reduction of the $\bar{d}-\bar{u}$ asymmetry
by the contribution from the $\pi \Delta$ Fock
states in the proton turns out much weaker than
evaluated before. 
We re-evaluated the GSR with the result
$\int_0^1 dx [\bar d - \bar u] \simeq  0.13 - 0.18$ and $S_G 
\simeq 0.21 - 0.25$.

As a spin off of our evaluation of isovector reggeon exchange 
to leading neutron production, we estimated which fraction of
the inclusive leading neutrons in DIS can be considered as due
to DIS off pions in the proton. In substantial part of the
phase space covered by the Forward/Leading Neutron Calorimeters installed
at HERA by H1 and ZEUS \cite{H1_neutron,ZEUS_neutron}, 
the purity of nuetron tagging of DIS
off pions can be estimated as 70-80 per cent. We feel that the
corresponding uncertainties with the determination of the pion 
structure function do not preclude useful test of its $x,Q^{2}$ 
evolution. To this effect we recall that judging from experimental
data on the proton structure function, the pion structure 
function is expected to vary by the factor $\sim 4$ from $x=10^{-1}$ 
to $x=10^{-4}$ and by the factor $\sim2$ from  $Q^2 =5$ to $Q^2 =100 \;GeV^2$ at 
$x = 3 \cdot 10^{-3}$ . 


\vskip 2cm

{\bf Acknowledgments:} \\
We are indebted to Y.V. Fisyak for the exchange of information
on the inclusive production of particles and G.T. Garvey for
the discussion of the E866 Fermilab data.
This work was supported partly by the German-Polish exchange program,
grant No. POL-81-97, and by the INTAS grant 96-0597.

\newpage

\newpage

\newpage

\newpage

\newpage

\newpage

\begin{center}
\begin{figure}
\epsfig{file=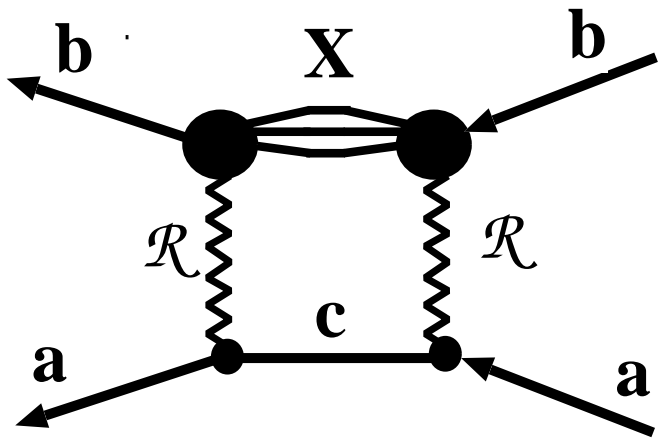, height = 6.0cm}
\caption{\it The amplitude for the double reggeon exchange.}
\label{total_diagram}
\end{figure}
\end{center}

\newpage

\newpage

\end{document}